\documentstyle[12pt]{article}
\newcommand{\ds}{\displaystyle}
\newcommand{\ba}{\begin{array}}
\newcommand{\ea}{\end{array}}
\newcommand{\be}{\begin{equation}}
\newcommand{\ee}{\end{equation}}
\newcommand{\bd}{\begin{displaymath}}
\newcommand{\ed}{\end{displaymath}}
\newcommand{\p}{\partial}

\newcommand{\ts}{\textstyle}
\newtheorem{theo}{Theorem}
\newtheorem{lem}{Lemma}
\newcommand{\bt}{\begin{theo}}
\newcommand{\et}{\end{theo}}
\newcommand{\nn}{\nonumber}
\title{Hermitian quasi-exactly solvable matrix Schr\"odinger operators}
\author{Stanislav Spichak\thanks{Electronic mail:
spichak@apmat.freenet.kiev.ua}\ \
and Renat Zhdanov\thanks{Electronic mail: rzhdanov@apmat.freenet.kiev.ua} \\
\small Institute of Mathematics, 3 Tereshchenkivska Street,
252004 Kyiv, Ukraine}
\date{}
\begin{document}
\maketitle

\begin{abstract}
We construct six multi-parameter families of Hermitian quasi-exactly
solvable matrix Schr\"odinger operators in one variable. The method for
finding these operators relies heavily upon a special representation of
the Lie algebra $o(2,2)\cong sl(2)\bigoplus sl(2)$ whose representation
space contains an invariant finite-dimensional subspace. Besides that we
give several examples of quasi-exactly solvable matrix models that have
square-integrable eigenfunctions. These examples are in direct analogy
with the quasi-exactly solvable scalar Schr\"odinger operators obtained
by Turbiner and Ushveridze.
\end{abstract}

\section*{I. Introduction}

In the paper \cite{zhd97a} we have suggested a generalization of the
Turbiner-Shifman approach \cite{tur88}--\cite{tus89} to the construction
of quasi-exactly solvable (QES) models on line for the case of matrix
Hamiltonians. We remind that originally their method was applied to
scalar one-dimensional stationary Schr\"odinger equations. Later on it
was extended to the case of multi-dimensional scalar stationary
Schr\"odinger equations \cite{tus89}--\cite{gko94b} (see, also
\cite{ush93}).

A systematic description of our approach can be found in the paper
\cite{zhd97b}. The procedure of constructing a QES matrix (scalar) model
is based on the concept of a Lie-algebraic Hamiltonian. We call a
second-order operator in one variable Lie-algebraic if the following
requirements are met:
\begin{itemize}

\item{The Hamiltonian is a quadratic form with constant coefficients of
first-order operators $Q_1, Q_2, \ldots, Q_n$ forming a Lie algebra
$g$;}

\item{The Lie algebra $g$ has a finite-dimensional invariant subspace
$\cal{I}$ of the whole representation space.}

\end{itemize}

Now if a given Hamiltonian $H[x]$ is Lie-algebraic, then after being
restricted to the space $\cal I$ it becomes a matrix operator $\cal H$
whose eigenvalues and eigenvectors are computed in a purely algebraic
way. This means that the Hamiltonian $H[x]$ is quasi-exactly solvable (for
further details on scalar QES models see \cite{ush93}).

It should be noted that there exist alternative approaches to
constructing matrix QES models \cite{tur92}--\cite{fgr97}. The principal
idea of these is fixing the form of basis elements of the invariant
space $\cal I$. They are chosen to be polynomials in $x$. This
assumption leads to a challenging problem of classification of
superalgebras by matrix-differential operators in one variable
\cite{fgr97}.

We impose no {\em a priori}\/ restrictions on the form of basis elements
of the space $\cal I$. What is fixed it is the class to which the basis
elements of the Lie algebra $g$ should belong. Following
\cite{zhd97a,zhd97b} we choose this class $\cal L$ as the set of matrix
differential operators of the form
\begin{equation} \label{0-1}
{\cal L} = \left\{ Q:\quad Q=a(x) \partial_x + A(x)\right\}.
\end{equation}
Here $a(x)$ is a smooth real-valued function and $A(x)$ is an $N\times
N$ matrix whose entries are smooth complex-valued functions of $x$.
Hereafter we denote $d/dx$ as $\partial_x$.

Evidently, $\cal L$ can be treated as an infinite-dimensional Lie
algebra with a standard commutator as a Lie bracket. Given a subalgebra
$\langle Q_1, Q_2, \ldots, Q_n \rangle $ of the algebra $\cal L$, that
has a finite-dimensional invariant space, we can easily construct a QES
matrix model. To this end we compose a bilinear combination of the
operators $Q_1, Q_2, \ldots, Q_n$ and of the unit $N\times N$ matrix $I$
with constant complex coefficients $\alpha_{jk}$ and get
\begin{equation}
\label{0-2}
H[x]\psi(x)=\left(\sum_{j,k=1}^n\alpha_{jk}Q_jQ_k\right).
\end{equation}

So there arises a natural problem of classification of subalgebras of
the algebra $\cal L$ within its inner automorphism group. The problem of
classification of inequivalent realizations of Lie algebras on line and
on plane has been solved in a full generality by Lie itself
\cite{lie24,lie27} (see, also \cite{gko92}). However, the classification
problem for the case when $A(x)\not = f(x)I$ with a scalar function
$f(x)$ is open by now. In the paper \cite{zhd97b} we have classified
realizations of the Lie algebras of the dimension up to three by the
operators belonging to $\cal L$ with an arbitrary $N$. Next, fixing
$N=2$ we have studied which of them give rise to QES matrix Hamiltonians
$H[x]$. It occurs that the only three-dimensional algebra that meets
this requirement is the algebra $sl(2)$ (which is fairly easy to predict
taking into account the scalar case!). This yields the two families of
$2\times 2$ QES models, one of them under proper restrictions giving
rise to the well-known family of scalar QES Hamiltonians (for more
details, see \cite{zhd97b}).

As is well-known a physically meaningful QES matrix Schr\"odinger
operator has to be Hermitian. This requirement imposes restrictions on
the choice of QES models which somehow were beyond considerations of our
previous papers \cite{zhd97a,zhd97b}. The principal aim of the present
paper is to formulate and implement an efficient algebraic procedure for
constructing QES Hermitian matrix Schr\"odinger operators
\begin{equation}
\label{0-3}
\hat H[x] = \partial_x^2 + V(x).
\end{equation}
This requires a slight modification of the algebraic procedure used in
\cite{zhd97b}. We consider as an algebra $g$ the direct sum of two
$sl(2)$ algebras which is equivalent to the algebra $o(2,2)$. The
necessary algebraic structures are introduced in Section 2. The next
Section is devoted to constructing in a regular way Hermitian QES matrix
Schr\"odinger operators on line. We give the list of thus obtained QES
models in Section 4. The fifth Section contains a number of examples of
Hermitian QES Schr\"odinger operators that have square integrable
eigenfunctions. 

\section*{II. Extension of the algebra $sl(2)$}

Following \cite{zhd97a,zhd97b} we consider the representation of the
algebra $sl(2)$
\be \label{11}
\ba{rcl}
sl(2)&=&\langle Q_-,\ Q_0,\ Q_+ \rangle \\[2mm]
&=&\langle \p_x,\ x\p_x -{{\ts m-1} \over {\ts 2}}+S_0,\
x^2\p_x -(m-1)x+2S_0x+S_+\rangle,
\ea
\ee
where $S_0=\sigma_3/ 2$,\ $S_+=(i\sigma_2+\sigma_1)/2$, $\sigma_k$ are
the $2\times 2$ Pauli matrices
$$
{\sigma_1}=\left(\begin{array}{cc}0 & 1\\ 1 &
0\end{array}\right),\
{\sigma_2}=\left(\begin{array}{cc}0 & -i\\ i &
0\end{array}\right),\
{\sigma_3}=\left(\begin{array}{cc}1 & 0\\ 0 &
-1\end{array}\right)
$$
and $m\geq 2$ is an arbitrary natural number. This representation gives
rise to a family of QES models and furthermore the algebra (\ref{11})
has the following finite-dimensional invariant space
\be \label{12}
\ba{l} {\cal I}_{sl(2)} ={\cal I}_1\bigoplus{\cal
I}_2 = \langle \vec e_1,x\vec e_1,\ldots,x^{m-2}\vec
e_1\rangle\bigoplus\\[2mm]
\langle m\vec e_2,\ldots,mx^j\vec e_2-jx^{j-1}\vec e_1,\ldots,
mx^m\vec e_2-mx^{m-1}\vec e_1\rangle.
\ea
\ee

Since the spaces ${\cal I}_1$, ${\cal I}_2$ are invariant with respect
to an action of any of the operators (\ref{11}), the above
representation is reducible. A more serious trouble is that it is not
possible to construct a QES operator, that is equivalent to a Hermitian
Schr\"odinger operator, by taking a bilinear combination (\ref{0-2}) of
operators (\ref{11}) with coefficients being complex numbers. To
overcome this difficulty we use the idea indicated in \cite{zhd97b} and
let the coefficients of the bilinear combination (\ref{0-2}) to be
constant $2\times 2$ matrices. To this end we introduce a wider Lie
algebra and add to the algebra (\ref{11}) the following three
matrix operators:
\be \label{13}
R_-=S_- ,\quad R_0=S_-x+S_0 ,\quad
R_+=S_-x^2+2S_0x+S_+ ,
\ee
where $S_{\pm}=(i\sigma_2\pm\sigma_1)/2$.

It is straightforward to verify that the space (\ref{12}) is invariant
with respect to an action of a linear combination of the operators
(\ref{13}). Consider next the following set of operators:
\be \label{14}
\langle T_{\pm}=Q_{\pm}-R_{\pm},\ T_0=Q_0-R_0,\ R_{\pm},\ R_0,\ I\rangle ,
\ee
where $Q$ and $R$ are operators (\ref{11}) and (\ref{13}), respectively,
and $I$ is a unit $2\times 2$ matrix. By a direct computation we check
that the operators $T_{\pm}, T_{0}$ as well as the operators $R_{\pm},
R_{0}$, fulfill the commutation relations of the algebra $sl(2)$.
Furthermore any of the operators $T_{\pm}, T_{0}$ commutes with any of
the operators $R_{\pm}, R_{0}$. Consequently, operators (\ref{14}) form
the Lie algebra
\[
sl(2)\bigoplus sl(2)\bigoplus I\cong o(2,2)\bigoplus I.
\]
In a sequel we denote this algebra as $g$.

The Casimir operators of the Lie algebra $g$ are multiples of the unit
matrix
$$
C_1=T_0^2-T_+T_--T_0=\left({{m^2-1} \over {4}}
\right)I ,\quad K_2=R_0^2-R_+R_--R_0={{3} \over {4}}I.
$$
Using this fact it can be shown that the representation of $g$ realized
on the space ${\cal I}_{sl(2)}$ is irreducible.

One more remark is that the operators (\ref{14}) satisfy
the following relations:
\be \label{15} \ba{lll}
R_-^2=0,\ \ R_0^2={{\ts 1} \over {\ts 4}},\ \ R_+^2=0, \\[2mm]
\{R_-,R_0\}=0,\ \ \{R_+,R_0\}=0,\ \ \{R_-,R_+\} =-1, \\[2mm]
R_-R_0={{\ts 1} \over {\ts 2}}R_-,\ \ R_0R_+={{\ts 1} \over {\ts 2}}R_+,\ \
R_-R_+=R_0-{{\ts 1} \over {\ts 2}}.  
\ea
\ee
Here $\{Q_1, Q_2\}=Q_1Q_2 + Q_2Q_1$. One of the consequences of this fact
is that the algebra $g$ may be considered as a superalgebra which shows
and evident link to the results of the paper \cite{fgr97}. 

\section*{III. The general form of the Hermitian QES operator}

Using the commutation relations of the Lie algebra $g$ together with
relations (\ref{15}) one can show that any bilinear combination of the
operators (\ref{14}) is a linear combination of twenty one (basis)
quadratic forms of the operators (\ref{14}). Composing this linear
combination yields all QES models which can be obtained with the help of
our approach. However the final goal of the paper is not to get some
families of QES matrix second-order operators as such but to get QES
Schr\"odinger operators (\ref{0-3}). This means that it is necessary to
transform bilinear combination (\ref{0-2}) to the standard form
(\ref{0-3}). What is more, it is essential that the corresponding
transformation should be given by explicit formulae, since we need to
write down explicitly the matrix potential $V(x)$ of thus obtained QES
Schr\"odinger operator and the basis functions of its invariant space.

The general form of QES model obtainable within the framework of our
approach is as follows
\be \label{6}
H[x]=\xi(x)\p_x^2+B(x)\p_x+C(x),
\ee
where $\xi(x)$ is some real-valued function and $B(x), C(x)$ are matrix
functions of the dimension $2\times 2$. Let $U(x)$ be an invertible
$2\times 2$ matrix-function satisfying the system of ordinary
differential equations
\be \label{8}
U'(x)={{\ts 1} \over {\ts 2\xi(x)}}\left({{\ts \xi'(x)} \over 
{\ts 2}}-B(x)\right)U(x),
\ee
and the function $f(x)$ be defined by the relation
\begin{equation}
\label{3-0}
f(x)=\pm\int {{\ts dx} \over {\sqrt{\xi(x)}}}.
\end{equation}
Then the change of variables reducing (\ref{6}) to the standard form
(\ref{0-3}) reads as
\begin{equation}
\label{7}
\begin{array} {rcl}
x&\rightarrow& y = f(x),\\[2mm]
H[x]&\rightarrow& \hat H[y] = \hat U^{-1}(y)H[f^{-1}(y)]\hat U(y),
\end{array}
\end{equation}
where $f^{-1}$ stands for the inverse of $f$ and $\hat U(y)=U(f^{-1}(y))$.

Performing the transformation (\ref{7}) yields the Schr\"odinger operator
\begin{equation}
\label{3-1}
\hat H[y]=\p_y^2+V(y)
\end{equation}
with
\be \label{9} \ba{rcl}
V(y)&=&\Biggl\{U^{-1}(x)\left[-{{\ts 1} \over {\ts 4\xi}}B^2(x)-{{\ts 1} 
\over {\ts 2}}B'(x)+{{\ts \xi'} \over {\ts 2\xi}}B(x)+C(x)\right]
U(x)\\[2mm]
&&\left.+{{\ts \xi''} \over {\ts 4}}-{{\ts 3{\xi'}^2}
\over {\ts 16\xi}}\Biggr\}\right|_{x=f^{-1}(y)} .
\ea
\ee
Hereafter, the notation $\{W(x)\}_{x=f^{-1}(y)}$ means that we should
replace $x$ with $f^{-1}(y)$ in the expression $W(x)$.

Furthermore, if we denote the basis elements of the invariant space
(\ref{12}) as $\vec f_1(x),\ldots, \vec f_{2m}(x)$, then the invariant
space of the operator $\hat H[y]$ takes the form
\be \label{10}
\hat {\cal I}_{sl(2)}=\left \langle \hat U^{-1}(y)\vec f_1(f^{-1}(y)),\ldots,
\hat U^{-1}(y)\vec f_{2m}(f^{-1}(y))\right \rangle .
\ee

In view of the remark made at the beginning of this section we are
looking for such QES models that the transformation law (\ref{7}) can be
given explicitly. This means that we should be able to construct a
solution of system (\ref{8}) in an explicit form. To achieve this goal
we select from the above mentioned set of twenty one linearly
independent quadratic forms of operators (\ref{14}) the twelve
forms,
\begin{eqnarray}
&&A_0=\p_x^2,\ A_1=x\p_x^2,\quad  A_2=x^2\p_x^2+(m-1)\sigma_3,\nn \\ 
&&B_0=\p_x,\ B_1=x\p_x+{{\ts \sigma_3} \over {\ts 2}},\quad 
B_2=x^2\p_x-(m-1)x+\sigma_3x+\sigma_1,\nn\\
&&C_1=\sigma_1\p_x+{{\ts m} \over {\ts 2}}\sigma_3,\quad 
C_2=i\sigma_2\p_x+{{\ts m} \over {\ts 2}}\sigma_3,\quad
C_3=\sigma_3\p_x,\label{16} \\ 
&&D_1=x^3\p_x^2-2\sigma_1x\p_x+(3m-m^2-3)x+(2m-3)x\sigma_3+(4m-4)\sigma_1,\nn\\
&&D_2=x^3\p_x^2-2i\sigma_2x\p_x+(3m-m^2-3)x+(2m-3)x\sigma_3+(4m-4)\sigma_1,\nn\\
&&D_3=2\sigma_3x\p_x+(1-2m)\sigma_3 ,\nn
\end{eqnarray}
whose linear combinations have such a structure that system (\ref{8}),
can be integrated in a closed form. However, in the present paper we
study systematically the first nine quadratic forms from the above list.
The quadratic forms $D_1, D_2, D_3$ are used to construct an example of
QES model such that the matrix potential is expressed via the
Weierstrass function.

Thus the general form of the Hamiltonian to be considered in a sequel
is as follows
\begin{eqnarray}
H[x]&=&\sum_{\mu=0}^2(\alpha_\mu A_\mu +\beta_\mu B_\mu)
+ \sum_{i=1}^3 \gamma_iC_i = (\alpha_2x^2+\alpha_1x+\alpha_0)\p_x^2\nn\\
&& +(\beta_2x^2+\beta_1x+\beta_0+
\gamma_1\sigma_1+i\gamma_2\sigma_2+\gamma_3\sigma_3)\p_x
+\beta_2\sigma_3x\label{17}\\
&&-\beta_2(m-1)x+\beta_2\sigma_1+\left[\alpha_2(m-1)+
{{\ts \beta_1} \over {\ts 2}}+{{\ts m} \over {\ts 2}}(\gamma_1+\gamma_2)
\right]\sigma_3.\nn
\end{eqnarray}
Here $\alpha_0,\alpha_1, \alpha_2$ are arbitrary real constants and $\beta_0,\ldots,
\gamma_3$ are arbitrary complex constants.

If we denote
\be\label{18}\ba{l}
\tilde\gamma_1=\gamma_1,\ \tilde\gamma_2=i\gamma_2,\
\tilde\gamma_3=\gamma_3,\ \delta =2\alpha_2(m-1)+
\beta_1+m(\gamma_1+\gamma_2),\\[2mm]
 \xi(x) =\alpha_2x^2+
\alpha_1x+\alpha_0,\ \eta(x) =\beta_2x^2+\beta_1x+\beta_0,
\ea
\ee
then the general solution of system (\ref{8}) reads as
\be\label{19}\ba{l}
\ds U(x)=\xi^{1/4}(x)\exp\left [-{{\ts 1} \over {\ts 2}}\int{{{\ts \eta(x)}
\over {\ts \xi(x)}}dx}\right ]\exp\left [-{{\ts 1} \over {\ts 2}}
\tilde\gamma_i\sigma_i\int{{{\ts 1}
\over {\ts \xi(x)}}dx}\right ]\Lambda ,
\ea
\ee
where $\Lambda$ is an arbitrary constant invertible $2\times 2$ matrix.
Performing the transformation (\ref{7}) with $U(x)$ being given by (\ref{19})
reduces QES operator (\ref{17}) to a Schr\"odinger form (\ref{3-1}),
where
\begin{eqnarray}
V(y)&=&\Biggl\{{{\ts 1} \over {\ts 4\xi}}\Lambda^{-1}\lbrace -\eta^2+2\xi 
'\eta-2\xi\eta '-4\beta_2(m-1)x\xi -\tilde\gamma_i^2\nn\\
&&+2(\xi '-\eta)\tilde\gamma_i\sigma_i+
4\beta_2\xi U^{-1}(x)\sigma_1U(x)+(4\beta_2x+2\delta)\xi\label{20} \\
&&\left.\times U^{-1}(x) \sigma_3U(x)\rbrace\Lambda+{{\ts \alpha_2} \over {\ts 2}}-
{{\ts 3(2\alpha_2x+\alpha_1)^2} \over {\ts 16\xi}}\Biggr\}
\right |_{x=f^{-1}(y)}.\nn
\end{eqnarray}
Here $\xi, \eta$ are functions of $x$ defined in (\ref{18}) and
$f^{-1}(y)$ is the inverse of $f(x)$ which is given by (\ref{3-0}).

The requirement of hermiticity of the Schr\"odinger operator (\ref{3-1})
is equivalent to the requirement of hermiticity of the matrix $V(y)$.
To select from the multi-parameter family of matrices (\ref{20}) Hermitian
ones we will make use of the following technical lemmas. 

\begin{lem} The matrices $z\sigma_a, w(\sigma_a\pm i\sigma_b), a\not
=b$,  with $\{z, w\}\subset{\bf C}, z\notin{\bf R}, w\not =0$ cannot be
reduced to Hermitian matrices with the help of a transformation
\begin{equation}
\label{3-2}
A\rightarrow A'=\Lambda^{-1}A\Lambda,
\end{equation}
where $\Lambda$ is an invertible constant $2\times 2$ matrix. 
\end{lem}
\vspace{2mm}

\noindent
{\bf Proof.}$\quad$ It is sufficient to prove the statement for the case
$a=1, b=2$, since all other cases are equivalent to this one. Suppose
the inverse, namely that there exists a transformation (\ref{3-2})
transforming the matrix $z\sigma_1$ to a Hermitian matrix $A'$. As ${\rm
tr}\ (z\sigma_1)={\rm tr}\ A'=0$, the matrix $A'$ has the form
$\alpha_i\sigma_i$ with some real constants $\alpha_i$. Next, from the
equality ${\rm det}\,(z\sigma_1)={\rm det}\,A'$ we get $z^2=\alpha_i^2$.
The last relation is in contradiction to the fact that $z\notin {\bf
R}$. Consequently, the matrix $z\sigma_1$ cannot be reduced to a
Hermitian matrix with the aid of a transformation (\ref{3-2}).

Let us turn now to the matrix $w(\sigma_1+i\sigma_2)$. Taking a general form
of the matrix $\Lambda$
\[
\Lambda=\left(\begin{array}{cc} a& b\\ c& d\end{array}\right)
\]
we represent (\ref{3-2}) as follows 
$$
A'=\Lambda^{-1}w(\sigma_1+i\sigma_2)\Lambda ={{\ts 2w} \over {\ts
\delta}}\left(\begin{array}{cc}cd & d^2\\ -c^2 & -cd
\end{array}\right),\quad \delta = {\rm det}\ \Lambda. 
$$
The conditions of hermiticity of the matrix $A'$ read
$$
{{\ts w} \over {\ts \delta}}cd={{\ts \bar w} \over {\ts
\bar\delta}}\bar c\bar d,\quad
{{\ts -w} \over {\ts \delta}}c^2={{\ts \bar w} \over {\ts
\bar\delta}}\bar d^2.
$$
where the bar over a symbol stands for the complex conjugation.

It follows from the second relation that $c, d$ can vanish only
simultaneously which is impossible in view of the fact that the matrix
$\Lambda$ is invertible. Consequently, the relation $cd\not = 0$ holds.
Hence we get
$$
{{\ts -d} \over {\ts c}}={{\ts \bar c} \over {\ts \bar d}}
\leftrightarrow |c|^2+|d|^2=0.
$$
This contradiction proves the fact that the matrix $w(\sigma_1 + i \sigma_2)$
cannot be reduced to a Hermitian form.

As the matrix $\sigma_1+i\sigma_2$ is transformed to become
$\sigma_1-i\sigma_2$ with the use of an appropriate transformation
(\ref{3-2}), the lemma is proved.

\begin{lem}
Let $\vec a=(a_1,a_2,a_3)$, $\vec b=(b_1,b_2,b_3)$, $\vec c=(c_1,c_2,c_3)$ be
complex vectors and $\vec \sigma$ be the vector whose components are the Pauli
matrices $(\sigma_1,\sigma_2,\sigma_3)$. Then the following assertions holds true.
\begin{enumerate}
\item{ A non-zero matrix $\vec a\vec\sigma$ is reduced to a Hermitian form with
the help of a transformation (\ref{3-2}) iff $\vec a^2>0$ (this
inequality means, in particular, that $\vec a^2\in{\bf R}$);}
\item{Non-zero matrices $\vec a\vec\sigma, \vec b\vec\sigma$ with $\vec b\not
=\lambda\vec a$,\ $\lambda\in{\bf R}$, are reduced simultaneously to Hermitian forms with
the help of a transformation (\ref{3-2}) iff
$$
\vec a^2>0,\ \ \vec b^2>0,\ \ (\vec a\times\vec b)^2>0;
$$}
\item{Matrices $\vec a\vec\sigma, \vec b\vec\sigma, \vec
c\vec\sigma$ with $\vec a\not =\vec 0,\ \vec b\not
=\lambda\vec a,\ \vec c\not
=\mu\vec b,\ \{\lambda,\mu\}\subset{\bf R}$ are reduced simultaneously to Hermitian forms with
the help of a transformation (\ref{3-2}) iff
\begin{eqnarray*}
&&\vec a^2>0,\quad \vec b^2>0,\quad (\vec a\times\vec b)^2> 0,\\
&& \left\{\vec a\vec c,\quad \vec b\vec c,\quad
(\vec a\times\vec b)\vec c\right\}
\subset {\bf R}. 
\end{eqnarray*}}
\end{enumerate}
Here we designate the scalar product of vectors $\vec a, \vec b$ as $\vec a\vec b$
and the vector product of these as $\vec a\times \vec b$.
\end{lem}
{\bf Proof.}$\quad$ Let us first prove the necessity of the assertion
1 of the lemma. Suppose that the non-zero matrix $\vec a \vec \sigma$
can be reduced to a Hermitian form. We will prove that hence it follows the
inequality $\vec a^2 > 0$.

Consider the matrices:
\be\label{21}\ba{l}
\Lambda_{ij}(a,b)=
\left\{ 
\ba{ll} 1+\epsilon_{ijk}{{\ts \sqrt{a^2+b^2}-b} \over 
{\ts a}}i\sigma_k, & a\not =0,\\[3mm] 1, & a=0, \ea
\right. 
\ea
\ee
where $(i,j,k)={\rm cycle}\ (1,2,3)$. It is not difficult to verify that
these matrices are invertible, provided
\be\label{22}
\sqrt{a_i^2+a_j^2}\not =0.
\ee

Given the condition (\ref{22}), the following relations hold
\be\label{23}
\sigma_l\rightarrow\Lambda^{-1}_{ij}(a,b)\ \sigma_l\
\Lambda_{ij}(a,b)=\left\{\begin{array}{ll} \sigma_k,&l=k,\\[2mm]
{{\ts b\sigma_i+a\sigma_j} \over {\ts \sqrt{a^2+b^2}}},&l=i,\\[6mm]
{{\ts -a\sigma_i+b\sigma_j} \over {\ts \sqrt{a^2+b^2}}}, &l=j.
\end{array}\right.
\ee

As $\vec a$ is a non-zero vector, there exists at least one pair of the
indices $i, j$ such that $a_i^2+a_j^2\not = 0$. Applying the
transformation (\ref{23}) with $a=a_i, b=a_j$ we get
\be \label{24}
\vec a\vec\sigma\rightarrow\vec a'\vec\sigma=
\sqrt{a_i^2+a_j^2}\, \sigma_j+a_k\sigma_k
\ee
(no summation over the indices $i,j,k$ is carried out). As the direct
check shows, the quantity $\vec a^2$ is invariant with respect to
transformation (\ref{23}), i.e. $\vec a^2= \vec a'^2$.

If $\vec a^2=0$, then ${a'}_j^2+{a'}_k^2=0$, or $a'_i=\pm ia'_k$. Hence by force of
Lemma 1 it follows that the matrix (\ref{24}) cannot be reduced to a Hermitian form.
Consequently, $\vec a^2\not =0$ and the relation  ${a'}_j^2+{a'}_k^2\not  =0$ holds
true. Applying transformation (\ref{23}) with $a=\sqrt{a_i^2+a_j^2},\ b=a_k$ we get
\be\label{25}
\vec a'\vec\sigma\rightarrow\sqrt{\vec a^2}\, \sigma_k.
\ee

Due to Lemma 1, if the number  $\sqrt{\vec a^2}$ is complex, then the
above matrix cannot be transformed to a Hermitian matrix. Consequently,
the relation $\vec a^2>0$ holds true.

The sufficiency of the assertion 1 of the lemma follows from the fact
that, given the condition $\vec a^2 > 0$, the matrix (\ref{25}) is
Hermitian.

Now we will prove the necessity of the assertion 2 of the lemma.
First of all we note that due to assertion 1, $\vec a^2>, \vec b^2
> 0$. Next, without loss of generality we can again suppose that
$a_i^2+a_j^2\not =0$. Taking the superposition of two transformations of
the form (\ref{23}) with $a=a_i, b=a_j$ and $a=\sqrt{a_i^2+a_j^2}, b=a_k$
yields
\be\label{26}\ba{l}
\Lambda_{ij}(a_i,a_j) \Lambda_{jk}(\sqrt{a_i^2+a_j^2},a_k)=
1+i\epsilon_{ijk}\frac{\ts \sqrt{\vec a^2}-a_k}{\ts \sqrt{a_i^2+
a_j^2}}\sigma_i\\[4mm]
\quad +i\epsilon_{ijk}\frac{\ts \sqrt{a_i^2+a_j^2}-
a_j}{\ts a_i}\sigma_k- i\epsilon_{ijk}\frac{\ts \sqrt{a_i^2+a_j^2}-a_j}{\ts a_i}
\frac{\ts \sqrt{\vec a^2}-a_k}{\ts \sqrt{a_i^2+a_j^2}}\sigma_j
\ea \ee
(here the finite limit exists when $a_i\rightarrow 0$).
Using this formula and taking into account (\ref{23}) yield
\be\label{27}
\vec a\vec\sigma\rightarrow\sqrt{\vec a^2}\sigma_k,\quad
\vec b\vec\sigma\rightarrow{\vec b'}\vec\sigma=
{{\ts b_ia_j-b_ja_i} \over {\ts  \sqrt{a_i^2+a_j^2}}}\sigma_i+
{{\ts a_k\vec a\vec b-b_k\vec a^2} \over
{\ts \sqrt{\vec a^2}\sqrt{a_i^2+a_j^2}}}\sigma_j+
{{\ts \vec a\vec b} \over  {\ts \sqrt{\vec a^2}}}\sigma_k.
\ee

Let us show that the necessary condition for the matrices $\sqrt{\vec
a^2}\sigma_k$,  $\vec b'\vec\sigma$ to be reducible to Hermitian forms
simultaneously reads as $\vec a\vec b\in{\bf R}$. Indeed, as the
matrices ${\vec b'}\vec\sigma, \sigma_k$ are simultaneously reduced to
Hermitian forms, the matrix $\vec b' \vec\sigma + \lambda \sigma_k$ can
be reduced to a Hermitian form with any real $\lambda$. Hence, in view
of the assertion 1 we conclude that
\be\label{28}
{b'}_i^2+{b'}_j^2+({b'}_k+\lambda)^2>0,
\ee
where $\lambda$ is an arbitrary real number. The above equality may be
valid only when $b'_k={\vec a\vec b\over \sqrt{\vec a^2}} \in {\bf R}$.

Choosing $\lambda = -b'_k$ in (\ref{28}) yields that
${b'}_i^2+{b'}_j^2>0$. Since ${b'}_i^2+{b'}_j^2=(\vec a\times\vec b)^2$,
hence we get the desired inequality $(\vec a\times\vec b)^2>0$. The necessity
is proved.

In order to prove the sufficiency of the assertion 2, we consider transformation
(\ref{23}) with
\be\label{29}
a={{\ts b_ia_j-b_ja_i} \over {\ts \sqrt{a_i^2+a_j^2}}},\quad
\ b={{\ts a_k\vec a\vec b-b_k\vec a^2} \over 
{\ts \sqrt{\vec a^2}\sqrt{a_i^2+a_j^2}}}.
\ee
This transformation leaves the matrix $\sqrt{\vec a^2}\sigma_k$ invariant, while
the matrix ${\vec b'}\vec\sigma$ (\ref{27}) transforms as follows
\be\label{30}
{\vec b'}\vec\sigma\rightarrow{\vec b''}\vec\sigma=
{{\ts \sqrt{(\vec a\times\vec b)^2}} \over {\ts
\sqrt{\vec a^2}}}\sigma_j+
{{\ts \vec a\vec b} \over 
{\ts \sqrt{\vec a^2}}}\sigma_k,
\ee
whence it follows the sufficiency of the assertion 2.

The proof of the assertion 3 of the lemma is similar to one of the assertion 2.
The first three conditions are obtained with account of the assertion 2.
A sequence of transformations (\ref{23}) with $a,b$ of the form (\ref{26}), (\ref{29})
transforms the matrix $\vec c\vec \sigma$ to become
\be\label{31}
\vec c\vec\sigma\rightarrow{\vec c\; ''}\vec\sigma=
{{\ts \epsilon_{ijk}\vec a(\vec c\times\vec b)} \over {\ts
\sqrt{(\vec c\times\vec b)^2}}}\sigma_i+
{{\ts (\vec a\times\vec b)(\vec a\times\vec c)} \over {\ts
\sqrt{(\vec c\times\vec b)^2}\sqrt{\vec a^2}}}\sigma_j+
{{\ts \vec a\vec c} \over 
{\ts \sqrt{\vec a^2}}}\sigma_k.
\ee
Using the standard identities for the mixed vector products we establish that
the coefficients by the matrices $\sigma_i, \sigma_j, \sigma_k$ are real
if and only if the relations
\[
\left\{\vec a\vec c,\ \vec b\vec c,\
(\vec a\times\vec b)\vec c\right\}
\subset {\bf R}
\]
hold true. This completes the proof of Lemma 2.

Lemma 2 plays the crucial role when reducing the potentials (\ref{20}),
to Hermitian forms. This is done as follows. Firstly, we reduce the QES operator
to the Schr\"odinger form 
\[
\partial_y^2 + f(y)\vec a\vec\sigma + g(y)\vec b\vec\sigma +
h(y)\vec c\vec\sigma + r(y).
\]
Here $f, g, h, r$ are some linearly-independent scalar functions and
$\vec a=(a_1,a_2,$ $a_3)$, $\vec b=(b_1,b_2,b_3)$, $\vec c=(c_1,c_2,c_3)$ are
complex constant vectors whose components depend on the parameters
$\vec\alpha, \vec\beta, \vec\gamma$. Next, using Lemma 2 we obtain the
conditions for the parameters $\vec\alpha, \vec\beta, \vec\gamma$ that
provide a simultaneous reducibility of the matrices $\vec a\vec\sigma,
\vec b\vec\sigma, \vec c\vec\sigma$ to Hermitian forms. Then, making use
of formulae (\ref{21}), (\ref{26}), (\ref{29}) we find the form of the
matrix $\Lambda$. Formulae (\ref{25}), (\ref{30}), (\ref{31}) yield
explicit forms of the transformed matrices $\vec a\vec\sigma, \vec
b\vec\sigma, \vec c\vec\sigma$ and, consequently, the Hermitian form of
the matrix potential $V(y)$.

\section*{IV. QES matrix models}

Applying the algorithm mentioned at the end of the previous section we
have obtained a complete description of QES matrix models (\ref{17})
that can be reduced to Hermitian Schr\"odinger matrix operators. We list
below the final result, namely, the restrictions on the choice of
parameters and the explicit forms of the QES Hermitian Schr\"odinger
operators and then consider in some detail a derivation of the
corresponding formulae for one of the six inequivalent cases. In the
formulae below we denote the disjunction of two statements $A$ and $B$
as $[A]\bigvee [B]$.
\vspace{2mm}

\noindent
{\bf Case 1.}\ $\tilde \gamma_1= \tilde \gamma_2=\tilde \gamma_3=0$ and
\[
\left[\beta_0,\ \beta_1,\ \beta_2\in{\bf R}\right]\bigvee 
[\beta_2=0,\ \beta_1=2\alpha_2,\
\beta_0=\alpha_1+i\mu ,\ \mu\in{\bf R}];
\]

\begin{eqnarray*}
\hat H[y]&=&\partial_y^2 + \Biggl\{{{\ts 1} \over
{\ts 4(\alpha_2x^2+\alpha_1x+\alpha_0)}}\lbrace 
-\beta_2^2x^4-[2\beta_1\beta_2+4\alpha_2\beta_2(m-1)]x^3\\
&&+\left[ 2\alpha_2\beta_1-2\alpha_1\beta_2-\beta_1^2-2\beta_0\beta_2-
4\alpha_1\beta_2(m-1)\right] x^2\\
&&+\left[ 4\alpha_2\beta_0-2\beta_0\beta_1-4m\alpha_0\beta_2\right] 
x+2\alpha_1\beta_0-2\alpha_0\beta_1-\beta_0^2\\
&&+4\beta_2(\alpha_2x^2+\alpha_1x+\alpha_0)\sigma_1
+(4\beta_2x+2\delta)(\alpha_2x^2+\alpha_1x+\alpha_0)\sigma_3
\rbrace\\
&&+{{\ts \alpha_2} \over {\ts 2}}-
{{\ts 3(2\alpha_2x+\alpha_1)^2} \over {\ts 
16(\alpha_2x^2+\alpha_1x+\alpha_0)}}\left.\Biggr\}\right|_{x=f^{-1}(y)},\\
\Lambda&=&1.
\end{eqnarray*}
{\bf Case 2.}\ $\beta_2, \delta=0$ and
\begin{eqnarray*}
&&2\alpha_2\beta_1-\beta_1^2\in{\bf R}, \
2\alpha_2\beta_0-\beta_0\beta_1\in{\bf R},\ 
2\alpha_1\beta_0-2\beta_1\alpha_0-\beta_0^2-\tilde\gamma_i^2\in{\bf R},\\
&&\left[(2\alpha_2-\beta_1)^2\tilde\gamma_i^2>0\right]\bigvee
\left[2\alpha_2-\beta_1=0\right],\
\left[(\alpha_1-\beta_0)^2\tilde\gamma_i^2>0\right]\bigvee
\left[\alpha_1-\beta_0=0\right];
\end{eqnarray*}
\begin{eqnarray*}
\hat H[y]&=&\partial_y^2 + \Biggl\{{{\ts 1} \over
{\ts 4(\alpha_2x^2+\alpha_1x+\alpha_0)}}\Biggl\{
\beta_1(2\alpha_2-\beta_1)x^2+2\beta_0(2\alpha_2-\beta_1)x\\
&&+2\alpha_1\beta_0-2\beta_1\alpha_0-\beta_0^2-\tilde\gamma_i^2+
\left[2(2\alpha_2-\beta_1)x+2(\alpha_1-\beta_0)\right]\sqrt{\tilde\gamma_i^2}
\sigma_3
\Biggr\}\\
&&+{{\ts \alpha_2} \over {\ts 2}}-
{{\ts 3(2\alpha_2x+\alpha_1)^2} \over 
{\ts 16(\alpha_2x^2+\alpha_1x+\alpha_0)}}\left.\Biggr\}\right|_{x=f^{-1}(y)},\\
\Lambda&=&\Lambda_{12}(\tilde\gamma_1,\tilde\gamma_2)
\Lambda_{23}(\sqrt{\tilde\gamma_1^2+\tilde\gamma_2^2},\tilde\gamma_3), \
\tilde\gamma_1^2+\tilde\gamma_2^2\not =0.
\end{eqnarray*}
(If $\tilde\gamma_1^2+\tilde\gamma_2^2=0$, then one can choose another matrix
$\Lambda$ (27) with $\tilde\gamma_i^2+\tilde\gamma_j^2\not =0$). \\
{\bf Case 3.}\ $\alpha_2\not = 0, \beta_2\not = 0$ and
\begin{eqnarray*}
&&\left[\{\beta_2, \gamma_1\}\subset {\bf{Re}},\ \gamma_3=0,\ \gamma_2 =
\sqrt{\gamma_1^2-2\alpha_2\gamma_1}, \alpha_2\gamma_1<0,\right.\\ 
&&\left. \beta_1=2\alpha_2+\beta_2{{\ts \alpha_1} \over {\ts \alpha_2}},\
\beta_0=\alpha_1+\beta_2{{\ts \alpha_0} \over {\ts \alpha_2}}\right];
\end{eqnarray*}
\begin{eqnarray*}
\hat H[y]&=& \partial_y^2 + \Biggl\{{{\ts \alpha_2} \over {\ts 2}}
-{{\ts 3(2\alpha_2x+\alpha_1)^2} 
\over {\ts 16(\alpha_2x^2+\alpha_1x+\alpha_0)}}
+{{\ts 1} \over {\ts 4(\alpha_2x^2+\alpha_1x+\alpha_0)}}
\Biggl\{-\beta_2^2x^4\\
&&-\left[2\beta_2^2{{\ts \alpha_1} \over {\ts \alpha_2}}+
4\alpha_2\beta_2m\right]x^3-\Biggl[{{\ts \beta_2^2} \over {\ts
\alpha_2^2}}(\alpha_1^2+2\alpha_0\alpha_2)
+2\alpha_1\beta_2(1+2m)\Biggr] x^2\\
&&-\left[ {{\ts 2\alpha_1\beta_2(\alpha_1\alpha_2+\alpha_0\beta_2)} 
\over {\ts \alpha_2^2}}+4\alpha_0\beta_2m\right] x
+\alpha_1^2-\beta_2^2{{\ts \alpha_0^2} \over {\ts
\alpha_2^2}}-4\beta_2{{\ts \alpha_0\alpha_1} \over {\ts \alpha_2}} -
4\alpha_0\alpha_2-2\alpha_2\gamma_1\\
&&+4\beta_2x(\alpha_2x^2+\alpha_1x+\alpha_0)\left[\sin\left(\theta(y)
\sqrt{-2\alpha_2 \gamma_1}\;\right)\sigma_1+
\cos\left(\theta(y)\sqrt{-2\alpha_2\gamma_1}\;\right)\sigma_3\right]\\
&&+2(\alpha_2x^2+\alpha_1x+\alpha_0)\Biggl[{{\ts \sin\left(\theta(y)
\sqrt{-2\alpha_2\gamma_1}\;\right)} \over {\ts
\sqrt{-2\alpha_2\gamma_1}}}\Biggl(\delta\sqrt{-2\alpha_2\gamma_1}\sigma_1
-2\beta_2\sqrt{\gamma_1^2-2\alpha_2\gamma_1}\sigma_3\Biggr)\\
&&+\left.\cos\left(\theta(y)\sqrt{-2\alpha_2\gamma_1}\;\right)\left(
{{\ts 2\beta_2\sqrt{\gamma_1^2-2\alpha_2\gamma_1}} \over {\ts 
\sqrt{-2\alpha_2\gamma_1}}}\sigma_1+
\delta\sigma_3\right)\Biggr]\Biggr\}\Biggr\}\right|_{x=f^{-1}(y)} ,\\
\Lambda&=&1+\left(\sqrt{1-{{\ts 2\alpha_2} \over {\ts \gamma_1}}}-
\sqrt{{{\ts -2\alpha_2} \over {\ts \gamma_1}}}\;\right)\sigma_3 .
\end{eqnarray*}
{\bf Case 4.}\ $\alpha_2\not =0, \beta_2 =0$.
\vspace{2mm}

\noindent
{\bf Subcase 4.1.}\ $\delta\not = 0$, $\gamma_1, \gamma_2$ do not vanish
simultaneously and
\[
\gamma_1^2-\gamma_2^2<0,\ \ \gamma_3=i\mu,\ \ \{\mu, \delta\}
\subset{\bf R},\ \ i(\alpha_1-\beta_0)\in{\bf R},\ \ \beta_1=2\alpha_2;
\]
\begin{eqnarray*}
\hat H[y]&=&\partial_y^2 + \Biggl\{{{\alpha_2} \over {2}}
-{{3(2\alpha_2x+\alpha_1)^2} 
\ds\over {16(\alpha_2x^2+\alpha_1x+\alpha_0)}}+{{1} \over {4\xi}}\Biggl\{
-\beta_0^2+2\alpha_1\beta_0-2\alpha_0
\beta_1-\tilde\gamma_i^2\\
&&+2(\alpha_2x^2+\alpha_1x+\alpha_0)\Biggl[\delta\sqrt{\gamma_2^2-\gamma_1^2}
\ds\sigma_1{{\sin\left(\theta(y)\sqrt{-\tilde\gamma_i^2}\;\right)} 
\ds \over {\sqrt{-\tilde\gamma_i^2}}}\\
&&+\ds {{-i\delta\gamma_3\sqrt{\gamma_2^2-\gamma_1^2}\sigma_2
\ds +\delta(\gamma_1^2-\gamma_2^2)\sigma_3} \over {\tilde\gamma_i^2}}
\cos\left(\theta(y)\sqrt{-\tilde\gamma_i^2}\;\right)\Biggr]
+\ds\Biggl[{{2\delta\alpha_2\gamma_3}\over{\tilde\gamma_i^2}}x^2\\
&&+\ds{{2\delta\alpha_1\gamma_3}\over{\tilde\gamma_i^2}}x+
\ds{{(2\alpha_1-2\beta_0)\tilde\gamma_i^2+2\delta\alpha_0\gamma_3}\over
\ds{\tilde\gamma_i^2}}\Biggr]
\left(i\sqrt{\gamma_2^2-\gamma_1^2}\sigma_2
+\gamma_3\sigma_3\right)\Biggr\}
\left.\Biggr\}\right|_{x=f^{-1}(y)},\\
\Lambda&=&\Lambda_{21}(i\gamma_1,\gamma_2).
\end{eqnarray*}

\noindent
{\bf Subcase 4.2.}\ $\delta\not =0,\ \gamma_1=\gamma_2=0,\ \gamma_3\not =0$ and
\begin{eqnarray*}
\{\delta ,\ \beta_1(2\alpha_2-\beta_1),\ \beta_0(2\alpha_2-\beta_1),\ 
-\beta_0^2+2\alpha_1\beta_0-2\alpha_0\beta_1,
\gamma_3(2\alpha_2-\beta_1),\ \gamma_3(\alpha_1-\beta_0)\}\subset {\bf R};
\end{eqnarray*}
\begin{eqnarray*}
\hat H[y]&=& \partial_y^2 + \Biggl\{{{\ts \alpha_2} \over {\ts 2}}-
{{\ts 3(2\alpha_2x+\alpha_1)^2} \over {\ts 16(\alpha_2x^2+\alpha_1x+\alpha_0)}}
+{{\ts 1} \over {\ts 4(\alpha_2x^2+\alpha_1x+\alpha_0)}}\lbrace
\beta_1(2\alpha_2-\beta_1)x^2\\
&&+2\beta_0 (2\alpha_2-\beta_1)x-\beta_0^2+
2\alpha_1\beta_0-2\beta_1\alpha_0-\gamma_3^2\\
&&+\left[2\delta\alpha_2x^2+2x((2\alpha_2-\beta_1)\gamma_3+\delta\alpha_1)+
2(\alpha_1-\beta_0)\gamma_3+2\delta\alpha_0\right]\sigma_3
\rbrace\left.\Biggr\}\right|_{x=f^{-1}(y)},\\
\Lambda&=&1 ,
\end{eqnarray*}
{\bf Case 5.}\ $\alpha_2=0,\ \beta_2\not =0$ and
\begin{eqnarray*}
&&\alpha_1\not =0,\ \gamma_1^2-\gamma_2^2<0,\ \tilde\gamma_i^2<0,\
\gamma_3= \ds{{\tilde\gamma_i^2}\over{2\alpha_1}},\\
&&\{\beta_0,\ \beta_1,\ \beta_2,\ \gamma_2,\ 
\delta(\gamma_2^2-\gamma_1^2)+2\beta_2\gamma_1\gamma_3\}\subset {\bf R},\\
&&\{i(2\alpha_0\beta_2\gamma_3-\beta_1\tilde\gamma_i^2+2\beta_2\alpha_1\gamma_1
+\delta\alpha_1\gamma_3),\ 
i((\alpha_1-\beta_0)\tilde\gamma_i^2+2\beta_2\alpha_0\gamma_1
+\delta\alpha_0\gamma_3)\} \subset {\bf R};
\end{eqnarray*}

\begin{eqnarray*}
\hat H[y]&=& \partial_y^2 + \Biggl\{- {{\ts 3\alpha_1^2} 
\over {\ts 16(\alpha_1x+\alpha_0)}}+
{{\ts 1} \over {\ts 4(\alpha_1x+\alpha_0)}}\Biggl\{
-\beta_2^2x^4-2\beta_1\beta_2x^3\\
&&+\left[(2-4m)\alpha_1\beta_2-\beta_1^2-2\beta_0\beta_2\right] x^2-
\left[ 2\beta_0\beta_1+4m\alpha_0\beta_2\right]x\\
&&+2\alpha_1\beta_0-2\alpha_0\beta_1-\beta_0^2-\tilde\gamma_i^2+
4x(\alpha_1x+\alpha_0)\Biggl[\beta_2\sqrt{\gamma_2^2-\gamma_1^2}\sigma_1
{{\ts \sin\left(\theta(y)\sqrt{-\tilde\gamma_i^2}\;\right)} \over {\ts 
\sqrt{-\tilde\gamma_i^2}}}\\
&&+{{\ts \beta_2\sqrt{(\gamma_1^2-\gamma_2^2)\tilde\gamma_i^2}} \over {\ts 
\tilde\gamma_i^2}}\sigma_3\cos\left(\theta(y)\sqrt{-\tilde\gamma_i^2}\;\right)
\Biggr]+2(\alpha_1x+\alpha_0)\\
&&\ds \times\Biggl[\left({{\delta(\gamma_2^2-\gamma_1^2)+2\beta_2\gamma_1\gamma_3} 
\ds \over {\sqrt{\gamma_2^2-\gamma_1^2}}}\sigma_1-
\ds {{2\beta_2\gamma_2\tilde\gamma_i^2} \over 
\ds {\sqrt{(\gamma_1^2-\gamma_2^2)\tilde\gamma_i^2}}}\sigma_3\right)
{{\ts \sin\left(\theta(y)\sqrt{-\tilde\gamma_i^2}\;\right)} \over {\ts 
\sqrt{-\tilde\gamma_i^2}}}\\
&&+\ds \left({{2\beta_2\gamma_2} \over 
\ds {\sqrt{\gamma_2^2-\gamma_1^2}}}\sigma_1+
\ds {{\delta(\gamma_1^2-\gamma_2^2)-2\beta_2\gamma_1\gamma_3} 
\ds \over
{\sqrt{(\gamma_1^2-\gamma_2^2)\tilde\gamma_i^2}}}\sigma_3\right) \cos\left(
\theta(y)\sqrt{-\tilde\gamma_i^2}\;\right)\Biggr]\\
&&+\Biggl[x{{\ts 4\alpha_0\beta_2\gamma_3-2\beta_1\tilde\gamma_i^2+
4\alpha_1\beta_2\gamma_1+2\delta\alpha_1\gamma_3} \over {\ts 
\tilde\gamma_i^2}}\\
&&+{{\ts (2\alpha_1-2\beta_0)\tilde\gamma_i^2+4\alpha_0\beta_2\gamma_1+2\delta
\alpha_0\gamma_3} \over {\ts 
\tilde\gamma_i^2}}\Biggr]\left(-i\sqrt{-\tilde\gamma_i^2}\sigma_2\right)
\Biggr\}\left.\Biggr\}\right|_{x=f^{-1}(y)},\\
\Lambda&=&\Lambda_{21}(i\gamma_1,\gamma_2)\Lambda_{23}\left(-i\gamma_3
\sqrt{\gamma_2^2-\gamma_1^2},\gamma_1^2-\gamma_2^2\right). 
\end{eqnarray*}
{\bf Case 6.}\ $\alpha_2=0,\ \beta_2=0$.
\vspace{2mm}

\noindent
{\bf Subcase 6.1.}\ $\delta\not =0$, $\tilde\gamma_1, \tilde\gamma_2$ do not vanish
simultaneously and 
\begin{eqnarray*}
&&\tilde\gamma_i^2<0,\ \{\delta^2(\gamma_1^2-\gamma_2^2)<0,\ \beta_0,\
\beta_1\} \subset {\bf R},\\
&&\{i(-\beta_1\tilde\gamma_i^2+\delta\alpha_1\gamma_3),\
i((\alpha_1-\beta_0)\tilde\gamma_i^2+\delta\alpha_0\gamma_3)\}\subset {\bf R};
\end{eqnarray*}

\begin{eqnarray*}
\hat H[y]&=& \partial_y^2 + \Biggl\{-{{\ts 3\alpha_1^2} 
\over {\ts 16(\alpha_1x+\alpha_0)}}
+{{\ts 1} \over {\ts 4(\alpha_1x+\alpha_0)}}\Biggl\{
-\beta_1^2x^2-2\beta_0\beta_1x+2\alpha_1\beta_0\\
&&-2\alpha_0\beta_1-\beta_0^2-\tilde\gamma_i^2
+\ds 2(\alpha_1x+\alpha_0)\Biggl[\delta\sqrt{\gamma_2^2-\gamma_1^2}\sigma_1
\ds {{\ts \sin\left(\theta(y)\sqrt{-\tilde\gamma_i^2}\;\right)} \over {\ts 
\ds \sqrt{-\tilde\gamma_i^2}}}\\
&&+\ds {{\delta(\gamma_1^2-\gamma_2^2)} \ds \over
\ds {\sqrt{(\gamma_1^2-\gamma_2^2)\tilde\gamma_i^2}}}\sigma_3\cos\left(
\theta(y)\sqrt{-\tilde\gamma_i^2}\;\right)\Biggr]+
\ds \Biggl[x{{\ts -2\beta_1\tilde\gamma_i^2+2\delta\alpha_1\gamma_3} \over {\ts 
\ds \tilde\gamma_i^2}}\\
&&+{{\ts (2\alpha_1-2\beta_0)\tilde\gamma_i^2+2\delta
\alpha_0\gamma_3} \over {\ts 
\tilde\gamma_i^2}}\Biggr]\left(-i\sqrt{-\tilde\gamma_i^2}\sigma_2\right)\Biggr\}
\left.\Biggr\}\right|_{x=f^{-1}(y)},\\
\Lambda&=&\Lambda_{21}(i\gamma_1,\gamma_2)\Lambda_{23}\left(-i\gamma_3
\sqrt{\gamma_2^2-\gamma_1^2},\gamma_1^2-\gamma_2^2\right).
\end{eqnarray*}
{\bf Subcase 6.2.}
\begin{eqnarray*}
&&\gamma_1=\gamma_2=0,\ \gamma_3\not =0,\ \{\beta_1^2,\
\beta_0\beta_1\}\subset {\bf R},\\ 
&&\{-\beta_1\gamma_3+\delta\alpha_1,\ (\alpha_1-\beta_0)\gamma_3
+\delta\alpha_0,\ -\beta_0^2+2\alpha_1\beta_0-2\alpha_0\beta_1\}
\subset {\bf R};
\end{eqnarray*}

\begin{eqnarray*}
\hat H[y]&=& \partial_y^2 + \Biggl\{-{{\ts 3\alpha_1^2} 
\over {\ts 16(\alpha_1x+\alpha_0)}}\\
&&+{{\ts 1} \over {\ts 4(\alpha_1x+\alpha_0)}}\lbrace
-\beta_1^2x^2-2\beta_0\beta_1x+
2\alpha_1\beta_0-2\alpha_0\beta_1-\beta_0^2-\gamma_3^2\\
&&+2(\alpha_1x+\alpha_0)[2x\beta_1(\alpha_1-\gamma_3)+2(\alpha_1-\beta_0)
\gamma_3+2\beta_1\alpha_0]\sigma_3
\rbrace\left.\Biggr\}\right|_{x=f^{-1}(y)},\\
\Lambda&=&1.
\end{eqnarray*}
In the above formulae we denote the inverse of the function
\be\label{34}
y=f(x)\equiv\int\ds {{dx}\over{\sqrt{\alpha_2x^2+\alpha_1x+\alpha_0}}},
\ee
as $f^{-1}(y)$ and, what is more, the function $\theta=\theta(y)$ is
defined as follows
\be \label{38}
\theta(y)=-\left.\left\{\int{{dx}\over{\alpha_2x^2+\alpha_1x+\alpha_0}}
\right\} \right|_{x=f^{-1}(y)},
\ee
and $\tilde\gamma_i^2$ stands for $\tilde\gamma_1^2 + \tilde \gamma_2^2
+ \tilde \gamma_3^2$.

The whole procedure of derivation of the above formulae is very
cumbersome. That is why we restrict ourselves to indicating the
principal steps of the derivation of the corresponding formulae for the
case when $\alpha_2\not =0, \beta_2\not =0$ omitting the secondary
details. It is not difficult to prove that $\tilde \gamma_i^2\not = 0$.
Indeed, suppose that the relation $\tilde \gamma_i^2 = 0$ holds and
consider the expression $\Omega =U^{-1}(x)\sigma_3U(x)$ from (\ref{20}).
Making use of the Campbell-Hausdorff formula we get
\[
\Omega =\sigma_3+\theta(i \gamma_1\sigma_2+\gamma_2\sigma_1)-
{{\ts \theta^2} \over {\ts 2}} \gamma_3\tilde\gamma_i\sigma_i,
\]
where $\theta$ is the function (\ref{38}). Considering the coefficient
at $\theta^2$, yields that $\gamma_3=0$ (otherwise using Lemma 2 we get
the inequality $\gamma_3^2\tilde\gamma_i^2\not = 0$ that contradicts to
the assumption $\tilde\gamma_i^2=0$). Since the matrix coefficient at
$\theta$ has to be Hermitian, we get
$\tilde\gamma_i^2=\gamma_1^2-\gamma_2^2<0$. This contradiction proves
that $\tilde\gamma_i^2\not =0$. Taking into account the proved fact we
represent the matrix potential (\ref{20}) as follows
\begin{eqnarray}
V(y)&=&\Biggl\{{{\ts \alpha_2} \over {\ts 2}}-{{\ts 3(2\alpha_2x+\alpha_1)^2} 
\over {\ts 16(\alpha_2x^2+\alpha_1x+\alpha_0)}}\nn\\
&&+{{\ts 1} \over {\ts 4(\alpha_2x^2+\alpha_1x+\alpha_0)}}\Lambda^{-1}\Biggl\{
-\beta_2^2x^4-[2\beta_1\beta_2+4\alpha_2\beta_2(m-1)]x^3\nn\\
&&+\left[ 2\alpha_2\beta_1-2\alpha_1\beta_2-\beta_1^2-2\beta_0\beta_2-
4\alpha_1\beta_2(m-1)\right] x^2\nn\\
&&+\left[ 4\alpha_2\beta_0-2\beta_0\beta_1-4m\alpha_0\beta_2\right] 
x+2\alpha_1\beta_0-2\alpha_0\beta_1-\beta_0^2-\tilde\gamma_i^2\nn\\
&&+4x(\alpha_2x^2+\alpha_1x+\alpha_0)\Biggl[\beta_2\gamma_3 
(\tilde\gamma_i^2)^{-1}
\tilde\gamma_i\sigma_i+\beta_2(\gamma_2\sigma_1+i\gamma_1\sigma_2)
(\tilde\gamma_i^2)^{-1/2}
\sinh\left(\theta\sqrt{\tilde\gamma_i^2}\;\right)\nn\\
&&+[\beta_2(-\gamma_1\gamma_3\sigma_1-i\gamma_2\gamma_3\sigma_2+
(\gamma_1^2-\gamma_2^2)\sigma_3)] 
(\tilde\gamma_i^2)^{-1}\cosh\left(\theta\sqrt{\tilde\gamma_i^2}\;\right)\Biggr]
\label{37}\\
&&+2(\alpha_2x^2+\alpha_1x+\alpha_0)\Biggl[(\delta\gamma_2\sigma_1+
i(\delta\gamma_1-2\beta_2\gamma_3)\sigma_2-
2\beta_2\gamma_2\sigma_3)(\tilde\gamma_i^2)^{-1/2}
\sinh\left(\theta\sqrt{\tilde\gamma_i^2}\;\right)\nn\\
&&+[(2\beta_2(\gamma_3^2-\gamma_2^2)-\delta\gamma_1\gamma_3)\sigma_1- 
i(2\beta_2\gamma_1\gamma_2+\delta\gamma_2\gamma_3)\sigma_2+(\delta
(\gamma_1^2-\gamma_2^2)-2\beta_2\gamma_1\gamma_3)\sigma_3]\nn\\
&&(\tilde\gamma_i^2)^{-1}\cosh\left(\theta\sqrt{\tilde\gamma_i^2}\;\right)
\Biggr]
+[(-2\beta_2\tilde\gamma_i^2+4\alpha_2\beta_2\gamma_1+2\delta
\alpha_2\gamma_3)x^2+
((4\alpha_2-2\beta_1)\tilde\gamma_i^2\nn\\
&&+4\alpha_1\beta_2\gamma_1+2\delta
\alpha_1\gamma_3)x
+{{\ts (2\alpha_1-2\beta_0)\tilde\gamma_i^2+4\alpha_0\beta_2\gamma_1+2\delta
\alpha_0\gamma_3} \over {\ts \tilde\gamma_i^2}}](\tilde\gamma_i^2)^{-1}\tilde\gamma_i
\sigma_i\Biggr\}\Lambda \left.\Biggr\}\right|_{x=f^{-1}(y)},\nn
\end{eqnarray}
where $\theta=\theta(y)$ is given by (\ref{38}).

Let us first suppose that $\gamma_1, \gamma_2$ do not vanish
simultaneously. We will prove that hence it follows that
$\tilde\gamma_i^2\in{\bf R}$. Consider the (non-zero) matrix coefficient
at $4x\xi\cosh\left(\theta\sqrt{\tilde\gamma_i^2}\;\right)$ in the
expression (\ref{37}) and suppose that $\sqrt{\tilde\gamma_i^2}=a+ib$,
with some non-zero real numbers $a$ and $b$. Now it is easy to prove that
$\cosh\left(\theta\sqrt{\tilde\gamma_i^2}\;\right)=f(x)+ig(x)$, where
$f, g$ are linearly-independent real-valued functions. Considering the
matrix coefficients of $f(x)$, $g(x)$ we see that in order  to reduce
the matrix (\ref{37}) to a Hermitian form we should reduce to Hermitian
forms the matrices $A$, $iA$ which is impossible. This contradiction
proves that $\tilde\gamma_i^2\in{\bf R}$.

Consider next the non-zero matrix coefficients of $4x\xi{{\ts
\sinh\left(\theta\sqrt{\tilde\gamma_i^2}\;\right)}  \over {\ts
\sqrt{\tilde\gamma_i^2}}},
4x\xi\cosh\left(\theta\sqrt{\tilde\gamma_i^2}\;\right)$ in (\ref{37}).
These coefficients can be represented in the form $\vec a \vec \sigma,\
\vec b \vec \sigma$, where
\[
\vec a=\beta_2(\gamma_2,i\gamma_1,0),\ \vec b=\beta_2(-\gamma_1\gamma_3,-i
\gamma_2\gamma_3,\gamma_1^2-\gamma_2^2),
\]
and, what is more,
\[
\vec a\times\vec b=\beta_2^2(\gamma_1^2-\gamma_2^2)(i\gamma_1,-\gamma_2,i
\gamma_3).
\]
Applying Lemma 2 yields
\[
\beta_i\in{\bf R},\ \ \gamma_3=i\mu ,\ \mu\in{\bf R},\ \
\gamma_1^2-\gamma_2^2<0.
\]

Next we turn to the matrix coefficient of $2\xi{{\ts \sinh\left(\theta
\sqrt{\tilde\gamma_i^2}\;\right)} \over {\ts \sqrt{\tilde\gamma_i^2}}}$ which is
of the form $\vec c\vec \sigma$ with $\vec
c=(\delta\gamma_2,i(\delta\gamma_1-
2\beta_2\gamma_3),-2\beta_2\gamma_2)$. Making use of the assertion 3
of Lemma 2 we obtain the conditions
\[
\{\gamma_1,\ \gamma_2\}\subset{\bf R},\ \ [\gamma_1=0]\bigvee[\gamma_3=0].
\]

Considering in a similar way the matrix coefficient of $2\xi\cosh\left(\theta
\sqrt{\tilde\gamma_i^2}\;\right)$ yields the following restrictions on
the coefficients $\vec \alpha, \vec \beta, \vec \gamma$:
\[
\Biggl[\{\beta_2,\gamma_1\}\subset{\bf{R}},\ \gamma_3=0, \gamma_2=
\sqrt{\gamma_1^2-2\alpha_2\gamma_1},\ \alpha_2\gamma_1<0,\\
\beta_1=2\alpha_2+\beta_2{{\ts \alpha_1} \over {\ts \alpha_2}},\
\beta_0=\alpha_1+\beta_2{{\ts \alpha_0} \over {\ts \alpha_2}}\Biggr].
\]
As a result we get the formulae of Case 2.

One can prove in an analogous way that, provided $\gamma_1=\gamma_2=0, \gamma_3\not =0$,
the matrix (\ref{37}) cannot be reduced to a Hermitian form.

\section*{V. Some examples.}

In this section we give several examples of Hermitian QES matrix Schr\"odinger
operators that have a comparatively simple form and, furthermore,  are
in direct analogy to QES scalar Schr\"odinger operators.
\vspace{2mm}

\noindent
{\em Example 1.}\ Let us consider Case I of the previous section with
$\alpha_0=\beta_2=1$, the remaining coefficients being equal to zero.
This choice of parameters yields the following Hermitian QES matrix
Schr\"odinger operator:
\be\label{58}
\hat H[y]=\p_y^2-{{\ts 1} \over {\ts 4}}y^4-my+\sigma_3y+\sigma_1.
\end{equation}
The invariant space ${\cal I}$ of the above Schr\"odinger
operator has the dimension $2m$ and is spanned by the
vectors
\[
\vec f_j=\exp \left({{\ts y^3} \over {\ts 6}}\right)\vec e_1y^j,\quad
\vec g_k=\exp \left({{\ts y^3} \over {\ts 6}}\right)(m\vec e_2y^k-k\vec 
e_yx^{k-1}),
\]
where $j=0,\ldots,m-2, k=0,\ldots, m$, $\vec e_1=(1,0)^T, \vec e_2=(0,1)^T$
and $m$ is an arbitrary natural number.

Note that the basis vectors of the invariant space $\cal I$ are
square-integrable on an interval $(-\infty, B]$ with an arbitrary
$B<+\infty$. It is also worth noting that there exists a QES scalar model
of the same structure that has analogous properties \cite{ush93}.

By construction, QES operator (\ref{58}) when restricted to the
invariant space $\cal I$ becomes complex $2m\times 2m$ matrix
$M$. However, the fact that operator (\ref{58}) is Hermitian does not
guarantee that the matrix $M$ will be Hermitian. It is straightforward
to check that the necessary and sufficient conditions of hermiticity of
the matrix $M$ read as
\begin{itemize}
\item{basis vectors $\vec f_j(y), \vec g_k(y)$ are square integrable
on the interval $[A, B]$,}
\item{the condition
\be\label{57}
(\p_y\vec r_j(y))\vec r_k(y)-\vec r_j(y)
(\p_y\vec r_k(y))|_A^B=0,
\ee
where $\vec r_i=f_i, i=0,\ldots, m-2$ and $r_i=\vec g_{i-m+1},
i=m-1,\ldots,2m-1$, holds $\forall j,k=0,\ldots,2m-1$.}
\end{itemize}
In the case considered relation (\ref{57}) does not hold and,
consequently, the matrix $M$ is not Hermitian. The next two examples are
free of this drawback, since the basis vectors of their invariant spaces
are square integrable on the interval $(-\infty, +\infty)$.
\vspace{2mm}

\noindent
{\em Example 2.}\ Let us now consider Case I of the previous section
with $\alpha_1=1, \beta_2=-1, \beta_0=1/2$, the remaining coefficients
being equal to zero. This choice yields the following QES matrix
Schr\"odinger operator
\[
\hat H[y] =\p_y^2 - {y^6\over 256} +{4m-1\over 16} y^2 -
{1\over 4}y^2 \sigma_3 -\sigma_1.
\]
The invariant space ${\cal I}$ of this operator has the dimension
$2m$ and is spanned by the vectors
\begin{eqnarray*}
\ds\vec f_j&=&\exp \left({{-y^4} \over {64}}\right)
\ds\left({{y}\over{2}}\right)^{2j}\vec e_1,\\
\ds\vec g_k&=&\exp \left({{-y^4} \over
{64}}\right)\left(m\left({{y}\over{2}}\right)^{2k}\vec e_2
\ds -k\left({{y}\over{2}}\right)^{2k-2}\vec e_1\right),
\end{eqnarray*}
where $j=0,\ldots, m-2,\ k=0,\ldots,m$.

It is not difficult to verify that the basis vectors of the invariant
space $\cal I$ are square integrable on the interval $(-\infty
,+\infty)$ and that the corresponding matrix $M$ is Hermitian.
One more remark is that there exists an analogous QES scalar
Schr\"odinger operator whose invariant space has square integrable
basis vectors (see, for more details \cite{tur88,ush88}).
\vspace{2mm}

\noindent
{\em Example 3.}\ Let us now consider Case III of the previous section
with $\alpha_2=1, \beta_2=-1, \gamma_1=-1$. This choice of parameters
yields the following QES matrix Schr\"odinger operator:
\begin{eqnarray*}
\hat H[y] &=& \p_y^2 -\ds{{1}\over{4}}-{{1}\over{4}}\exp(-2y)+
m\exp(-y)+{{1}\over{2}} \exp(2y)\\
&&+\ds \left[m{{\sqrt 3+1}\over{2}}\sin(\sqrt 2e^y)- \ds{{\sqrt
6}\over{2}}\cos(\sqrt 2e^y)-\exp(-y)\sin(\sqrt 2e^y)\right]\sigma_1\\
&&+\ds \left[m{{\sqrt 3+1}\over{2}}\cos(\sqrt 2e^y)+ \ds{{\sqrt
6}\over{2}}\sin(\sqrt 2e^y)-\exp(-y)\cos(\sqrt 2e^y)\right]\sigma_3.
\end{eqnarray*}
Furthermore, the invariant space ${\cal I}$ of this operator has the
dimension $2m$ and is spanned by the vectors
\begin{eqnarray*}
\vec f_j&=&U^{-1}(y)\exp(-jy)\vec e_1,\\
\vec g_k&=&U^{-1}(y)\left(m\exp(-ky)\vec e_2-
k\exp(-(k-1)y)\vec e_1\right),
\end{eqnarray*}
where $j=0,\ldots, m-2, k=0,\ldots, m$, $m$ is an arbitrary
natural number and
\begin{eqnarray*}
 U^{-1}(y)&=&{{1}\over{2\sqrt 2}}\exp\left(-{{y}\over{2}}\right)
\ds\exp\left(-{{1}\over{2}}e^{-y}\right)\\
&&\times (\sqrt 3+\sqrt 2-\sigma_3)\left[\cos(\sqrt 2e^y)+
\ds {{i\sqrt 3\sigma_2-\sigma_1}\over{\sqrt 2}}\sin(\sqrt
2e^y)\right].
\end{eqnarray*}

The basis vectors of the invariant space $\cal I$ are square
integrable and the condition (\ref{57}) holds. Indeed, the functions
$\vec f_j(y)$ and $\vec g_k(y)$ behave asymptotically as
$\exp\left(-{{(2j+1)y}\over{2}}\right)$ and
$\exp\left(-{{(2k+1)y}\over{2}}\right)$, correspondingly, with $y\to
+\infty$. Furthermore, they behave as $\exp\left(-{{(2j+1)y}
    \over{2}}\right)$ $\times \exp\left(-{{1}\over{2}}e^{-y}\right)$and
$\exp\left(-{{(2k+1)y}\over{2}}\right)
\exp\left(-{{1}\over{2}}e^{-y}\right)$, correspondingly, with $y\to
-\infty$. This means that they vanish rapidly provided $y\to \pm
\infty$.  \vspace{2mm}

\noindent
{\em Example 4.}\ The last example to be presented here is the QES
matrix model having a potential containing the Weierstrass function. To
this end we consider the whole set of operators (\ref{16}) and compose
the Hamiltonian
\[
H[x]=D_2 + A_1 +2B_2.
\]

Reducing $H[x]$ to the Schr\"odinger form yields the following QES
matrix model:
\begin{eqnarray*}
\hat H[y] &=& \p_y^2 + (m-m^2-1)w(y) -
{3(w(y)^2-1)^2\over 16(w(y)^3+w(y))}\\
&&+{2m-1\over w(y)^2+1}\left(2\sigma_1 + (w(y)^3+3w(y))\sigma_3\right).
\end{eqnarray*}
Here $m$ is an arbitrary natural number and $w(y)$ is the
Weierstrass function defined by the quadrature
\[
y=\int\limits_0^{w(y)}{dx\over \sqrt{x^3+x}}.
\]

The invariant space ${\cal I}$ of the operator $\hat H[y]$ has the
dimension $2m$ and is spanned by the vectors
\begin{eqnarray*}
\vec f_j&=&(w(y)^3 +w(y))^{-\frac{1}{4}}(1-i\sigma_2 w(y))
\exp(-jy)\vec e_1,\\
\vec g_k&=&(w(y)^3 +w(y))^{-\frac{1}{4}}(1-i\sigma_2 w(y))
\left(m\exp(-ky)\vec e_2\right.\\
&&\left.-k\exp(-(k-1)y)\vec e_1\right),
\end{eqnarray*}
where $j=0,\ldots, m-2, k=0,\ldots, m$.

Note that the first example of scalar QES model with elliptic
potential has been constructed by Ushveridze (see \cite{ush93}
and the references therein).

\section*{VI. Some conclusions}

A principal aim of the paper is to give a systematic algebraic treatment
of Hermitian QES Hamiltonians within the framework of the approach to
constructing QES matrix models suggested in our papers
\cite{zhd97a,zhd97b}. The whole procedure is based on a specific
representation of the algebra $o(2,2)$ given by formulae (\ref{11}),
(\ref{13}), (\ref{14}). Making use of the fact that the algebra
(\ref{14}) has an infinite-dimensional invariant subspace (\ref{12}) we
have constructed in a systematic way six multi-parameter families of
Hermitian QES Hamiltonians on line. Due to computational reasons we do
not present here a systematic description of Hermitian QES Hamiltonians
with potentials depending on elliptic functions (we give only an example
of such Hamiltonian in Section V).

The problem of constructing all Hermitian QES Hamiltonians of the form
(\ref{17}) having square-integrable eigenfunctions is also beyond the
scope of the present paper. We restricted our analysis of this problem
to giving two examples of such Hamiltonians postponing its further
investigation for our future publications.

A very interesting problem is a comparison of the results of the present
paper based on structure of representation space of the representation
(\ref{11}), (\ref{13}), (\ref{14}) of the Lie algebra $o(2,2)$ to those
of the paper \cite{fgr97}, where some superalgebras of
matrix-differential operators come into play. The link to the results of
\cite{fgr97} is provided by the fact that the algebra $o(2,2)$ has a
structure of a superalgebra. This is a consequence of the fact that
operators (\ref{14}) fulfill identities (\ref{15}).

One more challenging problem is a utilization of the obtained results
for integrating multi-dimensional Pauli equation with the help of the
method of separation of variables. As an intermediate problem to be
solved within the framework of the method in question is a reduction of
the Pauli equation to four second-order systems of ordinary differential
equations with the help of a separation Ansatz. The next step is
studying whether the corresponding matrix-differential operators belong
to one of the six classes of QES Hamiltonians constructed in Section IV.

Investigation of the above enumerated problems is in progress now and we
hope to report the results obtained in one of our future publications.

\section*{Acknowledgments}
This work is partially supported by the DFFD Foundation of Ukraine
(project 1.4/356).

\end{document}